\documentclass[useAMS,usenatbib]{mn2e}

\def\bea{\begin{eqnarray}}
\def\ena{\end{eqnarray}}
\title[Searching for GW in pulsars rotational parameters]{Investigating  ultra-long  gravitational waves with measurements of pulsars rotational parameters}
\author[M. S. Pshirkov ]{M. S. Pshirkov$^{1}$ \thanks{E-mail:pshirkov@prao.ru } \\
$^{1}$ PRAO ASC LPI, Pushchino, 142290, Russia}
\begin{document}

\date{}

\pagerange{\pageref{firstpage}--\pageref{lastpage}} \pubyear{2002}

\maketitle

\label{firstpage}

\begin{abstract}
A method is suggested to explore the gravitational wave background
(GWB) in the frequency range from $10^{-12}$ to
\hbox{$10^{-8}$\,Hz}. That method is based on the precise
measurements of pulsars' rotational parameters: the influence of
the gravitational waves (GW) in the range will affect them  and
therefore some conclusions about energy density of the GWB can be
made using   analysis of the derivatives of pulsars' rotational
frequency. The calculated values of the second derivative from a
number of pulsars limit the density of GWB $\Omega_{gw}$ as
follows: $\Omega_{gw}<2\times10^{-6}$.  Also, the time series of
the frequency $\nu$ of different pulsars in pulsar array can be
cross-correlated pairwise  in the same manner as in anomalous
residuals analysis thus providing the possibility of GWB detection
in ultra-low frequency range.

\end{abstract}


\section{Introduction}
Search for gravitational waves is one of the most important tasks
in  modern astronomy and physics. Various  techniques are used to
look for  gravitational waves in a very broad  range of
frequencies from $10^{-18}$ \citep{bgp2006} to \hbox{$10^{3}$\,Hz}
\citep{LIGOwebsite}.

Pulsar timing  provides a unique access  for observations  in a
low-frequency band ($10^{-7} ~\mathrm{Hz}<f_{gw}<10^{-9}
~\mathrm{Hz} $) \citep{Sazhin1978,Detweiler1979,Bertotti1983}.

Propagation of pulsar signal in space-time perturbed by a
stochastic gravitational wave field  results in apparent
deviations of pulsar rotational frequency; that influence can be
sought in anomalous residuals of pulses arrival times. RMS of that
residuals can be transferred to  the upper limits on density of
gravitational wave background (GWB). That background affects
signals from all pulsars, and common response can be extracted
from certain correlations in timing series
\citep{hd1983,Jenetetal2005,Anholmetal2008}, thus pulsar timing
can not only place upper limits, but can also  detect the presence
of gravitational waves as well. Like any other gravitational wave
detector, pulsar timing has its own frequency limitations. The
lowest frequency of the gravitational wave that can be observed
 corresponds to $T^{-1}_{\mathrm{obs}}$, where
$T_{\mathrm{obs}}$ -- total time span of pulsar observations
(usually years). Influence of gravitational waves of lower
frequencies simply redefines observed values of derivatives of
pulsar rotational parameters $\dot{\nu},\ddot{\nu}$. Sensitivity
of the method decreases with increase of GW frequency, so the
highest attainable frequency is of order of $10^{-7}~\mathrm{Hz}$.
Less stringent limitations  in the frequency region
$\sim(cD)^{-1}<f_{gw}< T^{-1}_{{\mathrm{obs}}}$, where $D$ --
distance to the pulsar, can be put using pulsar in binary system
as a precise clock. In that case, deviations from values of pulsar
orbital parameters predicted by General Relativity are treated as
manifestations of GWB induced effect and therefore some upper
limits on GWB density can be obtained
\citep{Bertotti1983,Kopeikin1997,Potapovetal2003}. That method
extends achievable frequency region down to
$10^{-12}~\mathrm{Hz}$. In this paper a similar  method is used:
gravitational waves affect the derivatives of pulsars rotational
frequency -- these values absorb the effect caused by
gravitational waves of ultra-low frequencies
($\sim(cD)^{-1}<f_{gw}<T^{-1}_{{\mathrm{obs}}}$). But the
information about gravitational wave background is not lost
completely and analysis of   rotational parameters can constrain
the characteristics of GWB in the frequency range from $10^{-12}$
to \hbox{$10^{-8}$\,Hz} \citep{Bertotti1983}. That method
complements the one that is based on observations of pulsar
binaries. Moreover, using time series of $\nu(t)$ in the same
manner as time series of ToA residuals \citep{Jenetetal2005} and
looking for the correlations between them we can try to detect GWB
in the ultra-low frequency region.

 Here we briefly review possible sources of GW in that frequency range. The gravitational radiation
from astrophysical sources like SMBH-binaries are negligible in
the region \citep{JaffeBacker2003}, so the main aim for the
searches is GWB of cosmological origin. There are a lot of
possible candidates for GW sources: relic gravitational waves from
primordial  fluctuations that were amplified during the
inflationary  stage in the Early Universe
\citep{Maggiore2000,glpps2001}, gravitational waves from phase
transitions in early Universe (e.g. \citep{Witten1984}),
gravitational waves from the string network \citep{dv2005}. The
spectrum  of the GWB depends on its sources, it can be either flat
(Harrison-Zeldovich spectrum) or it can possess different features
(e.g. peak at $\sim 10^{-12}~\mathrm{Hz}$ in some models of GWB
produced by strings \citep{VilenkinShellard1994}).

\section{Constraints from the second derivative of the rotation
frequency}
 Precise pulsar timing allows us to measure second
derivative of rotational frequency for a number of pulsars
\citep{hkl04}. However, that  value has a little in common with
physical value $\ddot{\nu}$ that somebody would expect from any
adopted laws of pulsar spin-down and usually surpasses the latter
by several orders of magnitude. So, it is very plausible to regard
the calculated value of $\ddot{\nu}$ as caused by unknown factors
that are  intrinsic or extrinsic to pulsar.

The technique of calculation in the paper follows \citep{bppp08}
with parameter $\epsilon=0$. We  work in the framework of a
slightly perturbed Minkowski space-time with coordinates $x^\mu =
(ct,x^i)$ and the metric given by:
\begin{equation}
 ds^2=-c^2dt^2+\left(\delta_{ij}+h_{ij}\right)dx^idx^j,
\label{metric}
\end{equation}
 where $h_{ij}$ is the gravitational
wave perturbation. Firstly, we  consider the simplest case of a
single monochromatic wave and then generalize it to the case of a
stochastic GWB.
 For a monochromatic gravitational wave the
metric perturbation $h_{ij}$ takes the form
\citep{LandauLifshitz,mtw}:
\begin{equation} h_{ij} = h~p_{ij}e^{ik_\mu x^\mu} =
h~p_{ij}e^{-i\left(k_0ct-k_ix^i\right)}, \label{singlegwmetric}
\end{equation} where $h$ is the amplitude of the gravitational wave, $k_\mu
= \left(k_0,k_i\right)$ is the wave vector, and $p_{ik}$ is the
polarization tensor of the gravitational wave.  We can introduce a
set of two mutually orthogonal unit vectors $l_i$ and $m_i$
orthogonal to the wave vector $k_i$;   the polarization tensor
$p_{ik}$ has the form \citep{LandauLifshitz,mtw}.
 \bea
p_{ik} = \frac{1}{2}\left( l_i\pm m_i \right)\left( l_k\pm m_k
\right), \label{defpolten} \ena where $\pm$ corresponds to  two
independent states of circular polarization. Due to the transverse
and traceless nature of gravitational waves, the polarization
tensor satisfies the following conditions \bea p_{ik}k^i = 0,
~~~p_{ik}\delta^{ik} = 0. \label{TTconditions} \ena It is
convenient to introduce the wavenumber
$k=\left(\delta_{ij}k^ik^j\right)^{1/2}$,  and a unit vector in
the direction of wave propagation $\tilde{k}^i=k^i/k$. The
wavelength of the gravitational wave is related to the wavenumber
by the equality $k = 2\pi/\lambda_{gw}$ and  the frequency of the
gravitational wave ${f_{gw}}$ is related to the time component of
the wave vector through the relation $k_0=2{\pi}{f_{gw}}/c$.

The effect of a gravitational wave upon the measured frequency of
pulsar signal was considered in \citep{Sazhin1978,Detweiler1979}.

We can write down the final result:  \bea
\frac{\delta\nu(t)}{\nu_0} =
\frac{1}{2}h~e^ie^jp_{ij}~e^{-ikct}~\left[
\frac{1-e^{i\left(1-\tilde{k}_ie^i\right)kD}}{\left(1-\tilde{k}_ie^i\right)}
\right] \label{deltanu2} \ena where $\nu_0$ is the unperturbed
pulsar frequency in the absence of gravitational waves and
$\delta\nu(t) = \nu(t) - \nu_0$ is the variation of pulsar
frequency due to their presence. $D$ is the distance from the
pulsar to the observer, $e^i$ is the unit vector tangent along
this path (i.e.~unit vector in the direction from pulsar to the
observer).

Gravitational wave will  also affect  derivatives of pulsar
frequency analogously:
 \bea
\frac{\delta\dot{\nu}(t)}{\nu_0} =
\frac{-ikc}{2}h~e^ie^jp_{ij}~e^{-ikct}~\left[
\frac{1-e^{i\left(1-\tilde{k}_ie^i\right)kD}}{\left(1-\tilde{k}_ie^i\right)}
\right]\label{deltader1nu} \ena

 \bea
  \frac{\delta\ddot{\nu}(t)}{\nu_0} =
\frac{-k^2c^2}{2}h~e^ie^jp_{ij}~e^{-ikct}~\left[
\frac{1-e^{i\left(1-\tilde{k}_ie^i\right)kD}}{\left(1-\tilde{k}_ie^i\right)}
\right]. \label{deltader2nu} \ena

Our analysis can be generalized to a stochastic  gravitational
wave background case. That background  can be decomposed into
spatial Fourier modes: \bea h_{ij}(t,x^i) = \int d^3{\bmath{k}}
\sum_{s=1,2} \left[ h_s(k^i,t)\stackrel{s}{p}_{ij}(k^l)e^{ik_ix^i}
+ c.c. \right], \label{fouriergw} \ena where $d^3{\bmath k}$
denotes the integration over all possible wave vectors, "c.c"
stands for "complex conjugate" and $s=1,2$ corresponds to  two
linearly independent modes of polarization satisfying the
orthogonality condition \bea
\stackrel{s}{p}_{ij}\stackrel{s'}{p}{}^{ij*} = \delta_{ss'}
\label{poltenorthog} \ena The mode function $h_s(k^i,t)$
corresponds to plane monochromatic waves \bea h_s(k^i,t) =
h_s(k^i)~e^{-ikct} \label{hmodefunctions} \ena Due to the
linearity of the equations,  shifts in derivatives of frequency
can be presented as integrals:
 \bea \frac{\delta\ \frac{d^{1,2}\nu}{dt^{1,2}}}{\nu_0} = \int d^3{\bmath{k}}
\sum_{s=1,2} \left[ h_s(k^i)\tilde{R}_{+1,+2}(t;k^i,s) +
c.c.\right] \label{fourierR}, \ena where indices $+1,+2$ refer to
the first and the second derivative respectively. Applying  the
results from the consideration of a single monochromatic wave, we
can write down the contribution from a single Fourier component
$\tilde{R}_{+1,+2}(t;k^i,s)$:
 \bea
\tilde{R}_{+1}(t;k^i,s) =
\frac{-ikc}{2}~e^ie^jp_{ij}~e^{-ikct}~\left[
\frac{1-e^{i\left(1-\tilde{k}_ie^i\right)kD}}{\left(1-\tilde{k}_ie^i\right)}\right],
\label{trans1_1}
 \ena

 \bea
\tilde{R}_{+2}(t;k^i,s) =
\frac{-k^2c^2}{2}~e^ie^jp_{ij}~e^{-ikct}~\left[
\frac{1-e^{i\left(1-\tilde{k}_ie^i\right)kD}}{\left(1-\tilde{k}_ie^i\right)}\right],
\label{trans1_2}
 \ena

 where the tilde over $R_{+1,+2}$ in the above
expressions is introduced to indicate explicit factoring out of
the gravitational wave amplitude $h$.

Usually we know only statistical properties of the gravitational
wave field. Stationary statistically homogeneous and isotropic
gravitational wave field possesses the following properties: \bea
<h_s(k^i)> = 0,\\ <h_s(k^i)~ h_{s'}^{*}(k'^i)> =
\frac{P_h(k)}{16\pi k^3}\delta_{ss'}\delta^3(k^i-k'^i),
\label{gwstatprop} \ena where the brackets denote ensemble
averaging over all possible realizations, and $P_h(k)$ is the
metric power spectrum per logarithmic interval of $k$.  Using
(\ref{gwstatprop}), we can  calculate the statistical properties
of the corresponding shifts in frequency derivatives
$\dot{\nu}(t)$ and  $\ddot{\nu}(t)$. Using (\ref{fourierR}) and
(\ref{gwstatprop}), and taking into account the orthogonality
property (\ref{poltenorthog}), after straight forward
calculations, we arrive at the following statistical expressions:

\bea <\frac{\delta\dot{\nu}(t)}{\nu_0}> &=& 0,
\label{nudotmean}\\
 <\left(\frac{\delta\dot{\nu}(t)}{\nu_0}\right)^2>
&=& \int \frac{dk}{k} P_h(k) \tilde{R}_{+1}^2(k),
\label{nudotsquaremean} \\
<\frac{\delta\ddot{\nu}(t)}{\nu_0}> &=& 0,
\label{nuddotmean}\\
 <\left(\frac{\delta\ddot{\nu}(t)}{\nu_0}\right)^2>
&=& \int \frac{dk}{k} P_h(k) \tilde{R}_{+2}^2(k),
\label{nuddotsquaremean} \ena

 where we have introduced the transfer functions
\bea \tilde{R}_{+1,2}^2(k) = \frac{1}{8\pi}\int d\Omega \sum_s
\left| \tilde{R}_{+1,2}(t;k^i,s) \right|^2.
\label{transferfunction} \ena In the above expression $d\Omega$
represents integration over the possible directions of
gravitational wave (i.e.~$d^3{\bmath{k}} = k^2dkd\Omega$). From
(\ref{trans1_1},\ref{trans1_2}) and (\ref{transferfunction}) it
follows that the transfer functions $\tilde{R}_{+1,+2}^2(k)$ do
not depend on time variable $t$. That results from the
stationarity of the gravitational wave field.

The expressions for the transfer function are  calculated in the
similar  way as in \citep{bppp08}, but now they are slightly more
complicated:

\bea \tilde{R}_{+1}^2(k) \approx
\frac{k^2c^2}{6}-\frac{c^2}{4D^2}+\frac{\cos{(kD)}\sin{(kD)}c^2}{4kD^3}
\label{transferfunction-1} \ena \bea \tilde{R}_{+2}^2(k) \approx
\frac{k^4c^4}{6}-\frac{k^2c^4}{4D^2}+\frac{\cos{(kD)}\sin{(kD)}kc^4}{4D^3}.
\label{transferfunction-2} \ena

These transfer functions behave like $k^4$  and $k^6$ respectively
when $k\rightarrow0$.

The statistical properties of stochastic gravitational wave field
may be characterized by the density parameter $\Omega_{gw}$
\citep{Allen1997}. $\Omega_{gw}$ is related to the power spectrum
$P_h(k)$:
 \bea \Omega_{gw}(k)  =   \frac{2\pi^2}{3}
\left( \frac{k}{k_H}\right)^2P_h(k)
 \label{definitionofOmega}
\ena where $k_H= 2\pi f_H/c = 2\pi H_0/c$, and $H_0$ is the
current Hubble parameter. The density parameter $\Omega_{gw}$ is
the current day ratio of energy density of gravitational waves
(per unit logarithmic interval in $k$) to the critical density of
the Universe $\rho_{crit} = 3c^2H_0^2/8\pi G$. For numerical
estimations, we set Hubble parameter $H_0 = 75~\frac{{\rm
km}}{{\rm sec}}/{\rm Mpc}$ and assume a simple  power law spectrum
for the density parameter $\Omega_{gw}$: \bea \Omega_{gw} (k)=
\Omega_{gw} (k_o) \left(\frac{k}{k_o}\right)^{n_T}.
\label{powerlawspectrum} \ena

This form of spectrum can be used as a good approximation for a
large variety of models in frequency range of our interest.  The
flat, scale invariant power spectrum (also known as
Harrison-Zeldovich power spectrum) corresponds to $n_T=0$. To
obtain mean square deviations, we should integrate
(\ref{nudotsquaremean}) and (\ref{nuddotsquaremean}) with GW
spectrum (\ref{powerlawspectrum}). The limits of integration
$k_{\mathrm{min}}$ and $k_{\mathrm{max}}$ are determined from
following considerations: the highest frequency that puts in  the
effect is defined by cut-off scale coming from the pulsar timing
technique and determined by total time span of pulsar observations
used to obtain rotational parameters, the lowest frequency of GW
that can be probed with that method comes from the limitations of
pulsar-Earth distance\footnote{The method can probe cosmological
GW with length up to present-day horizon size, but effectively
their contribution is suppressed by transfer functions that
approach 0 quickly; they should only be taken into  account in
case of  GWB spectrum  that reddens towards lower frequencies.};
$k_{\mathrm{min}}\approx\frac{2\pi}{D}$ and
$k_{\mathrm{max}}=2{\pi}/cT_{{\mathrm{obs}}}$ (note that our
maximal frequency coincides with minimal frequency of usual pulsar
timing searches for GWs; in fact
$k_{\mathrm{max}}=2\alpha{\pi}/cT_{{\mathrm{obs}}},~\alpha\simeq
1$, for the sake of simplicity we used $\alpha=1$). After
substitutions we obtain the following equations: \bea
<\left(\frac{\delta\dot{\nu}(t)}{\nu_0}\right)^2> =
\frac{c^2}{4\pi^2}\Omega_{gw}(k_0)k_H^2\int k^{n_T-1}=
\label{nudotsquaremean3} \ena
$$= \left\{
\begin{tabular}{l}
$\frac{c^2}{4\pi^2 n_T}\Omega_{gw}(k_0)k_H^2k_0^{-n_T}[k_{\mathrm{max}}^{n_T}-k_{\mathrm{min}}^{n_T}],~n_T\neq0$\\$\frac{c^2}{4\pi^2 }\Omega_{gw}k_H^2 \ln{\left(\frac{k_{\mathrm{max}}}{k_{\mathrm{min}}}\right)},~n_T=0$\\
\end{tabular} \right.
$$
and

\bea <\left(\frac{\delta\ddot{\nu}(t)}{\nu_0}\right)^2> =
\frac{c^4}{4\pi^2}\Omega_{gw}(k_0)k_H^2\int k^{n_T+1}=
\label{nuddotsquaremean3} \ena
$$= \left\{
\begin{tabular}{l}
$\frac{c^4}{4\pi^2(n_T+2)}\Omega_{gw}(k_0)k_H^2k_0^{-n_T}[k_{\mathrm{max}}^{n_T+2}-k_{\mathrm{min}}^{n_T+2}],~n_T\neq-2$\\$
\frac{c^4}{4\pi^2}\Omega_{gw}(k_0)k_0^2k_H^2\ln{\left(\frac{k_{\mathrm{max}}}{k_{\mathrm{min}}}\right)},~n_T=2$,\\
\end{tabular} \right.
$$

where two last terms in eqns.
(\ref{transferfunction-1},\ref{transferfunction-2}) are omitted
because they are always much smaller than the first one.
 More stringent constraints come from expression
({\ref{nuddotsquaremean3}}) because the GW influence on the first
derivative is hidden inside spin-down part.

As observed value of the second derivative
$\ddot{\nu}_{{\mathrm{obs}}}$ can not exceed the  value produced
by the effect we can substitute
$<\left(\frac{\delta\ddot{\nu}(t)}{\nu_0}\right)^2>$ with
$\left(\frac{\ddot{\nu}_{{\mathrm{obs}}}}{\nu_0}\right)^2$ in
({\ref{nuddotsquaremean3}})\footnote{As a matter of fact, second
derivative in pulsar timing analysis results from fitting a cubic
polynomial  to observational data. Obtained value slightly differs
from actual second derivative of pulsar residuals owing to the
contributions to fitting procedure from higher than third order
terms and noise terms; I am grateful to referee for pointing that
out.}. Finally, we arrive at the following constraint (adopting
the flat spectrum of GW):

\begin{equation}
\label{finalomega1}
\Omega_{gw}<\frac{8\pi^2}{c^4k_H^2k_{\mathrm{max}}^2}\left(\frac{\ddot{\nu}_{{\mathrm{obs}}}}{\nu_0}\right)^2,
\end{equation}
or
\begin{equation}
\label{finalomega2}
\Omega_{gw}<\frac{T_{{\mathrm{obs}}}^2}{2\pi^2H_0^2}\left(\frac{\ddot{\nu}_{{\mathrm{obs}}}}{\nu_0}\right)^2
\end{equation}

It is instructive to make some numerical estimates using the data
for PSR B1937+21; that pulsar has been timed for a long time with
a very high precision. The values of rotational parameters is
taken from \citep{manchester}: \hbox{$\nu_0=641$\,Hz},
\hbox{$\frac{\ddot{\nu}}{\nu_0}=6.2\cdot10^{-29}$\,s$^{-2}$}
\begin{equation}
\label{1937limits}
\Omega_{gw}<2.8\times10^{-6}\left(\frac{T_{{\mathrm{obs}}}}{10~yrs}\right)^2
\end{equation}
these constraints are 2 orders of magnitude stronger than the
current limit \citep{Kopeikin1997}. It is intriguing that
virtually all  pulsars with small  second derivatives exhibit
approximately the same value of defining relation
$10^{-29}<\left|\frac{\ddot{\nu}}{\nu_0}\right|<10^{-28}$\,s$^{-2}$
\footnote{Except PSR J1952+1410
($\frac{\ddot{\nu}}{\nu_0}=-6.6\cdot10^{-30}$\,s$^{-2}$),
J1946+1805 ($\frac{\ddot{\nu}}{\nu_0}=8.8\cdot10^{-30}$\,s$^{-2}$)
and J1823+0550
($\frac{\ddot{\nu}}{\nu_0}=-5.1\cdot10^{-30}$\,s$^{-2}$). Placing
that values into (\ref{finalomega2}) produces even more stringent
limits on $\Omega_{gw}$.}. That relation holds both for ordinary
and millisecond pulsars. The part in $\ddot{\nu}$ that is induced
by the effect, has a $T_{{\mathrm{obs}}}^{-1}$ dependance   on
total observation time span. Having long  high-quality timing
series we would be able to obtain function
$\ddot{\nu}_{{\mathrm{obs}}}(T_{{\mathrm{obs}}})$ and then extract
from it the part that can be caused by GW influence , thus
effectively making limits on $\Omega_{gw}$ tighter.

\section{Possible method of detection of ultra-low frequency GW}\
The technique proposed in the previous section can only place
limits on the existence of ultra-low frequency GWB; it can not be
used to detect the presence of that background. However, having
 time series  of some parameter that is affected by the GWB for several pulsars, we can use pairwise
cross-correlations in the same way they are used in residuals
analysis \citep{hd1983,Jenetetal2005}. Actually, the time series
for $\nu(t)$ can be used instead of $r(t)$ in "classical" case
\citep{ktr1994, Jenetetal2006}. Data are prepared as follows:
total span of  timing observations is sampled into shorter
sub-intervals, e.g. one year long and   frequency is calculated
for every interval. Magnitude of frequency will decrease linearly
due to the  regular spin-down effect. The first-order contribution
from the GWB would be absorbed by that much more substantial
effect. However, the GWB induced effect can be sought in the
series of observed frequencies after removing of linear trend. The
correlation coefficient between the observed values of $\Delta\nu$
for each pair of pulsar has the form:
\begin{equation}
\label{correl1}
 f(\theta)=\frac{1}{N}\sum_{i=0}^{i=N-1}\frac{\Delta\nu_1(t_i,\bmath{e_1})}{\nu_{01}}
 \frac{\Delta\nu_{2}(t_i,\bmath{e_2})}{\nu_{02}},
\end{equation}
where $N$ -- number of sub-intervals, $\bmath{e_1},\bmath{e_2}$ --
unit vector in the directions to pulsars, $\nu_{0i}$ -- average
values of pulsars rotational frequencies.

Due to the linearity, differentiation with respect to time does
not affect the angular dependence of correlation function on the
angular distance between two pulsars  $\theta$
($\cos{\theta}=\bmath{e_1}\cdot\bmath{e_2}$) and:
\begin{equation}
\label{correl2}
 <f(\theta)>=\sigma_{\Delta\nu}^2\zeta(\theta),
 \end{equation}
 $$\zeta(\theta)=\frac{3(1-\cos{\theta})}{4}\log{\frac{1-\cos{\theta}}{2}}-\frac{1-\cos{\theta}}{8}+\frac{1}{2}+\frac{1}{2}\delta(\theta),$$
where $\delta{\theta}$ equals 1 when $\theta=0$ and 0 otherwise.
Correlated part  $\Delta\nu$ allows us to detect  presence of GWB,
$\sigma_{\Delta\nu}^2=<\left(\frac{\Delta\nu}{\nu_0}\right)^2>$.

Correlated part $\Delta \nu$ induced by GWB can be very roughly
estimated as:
\begin{equation}
\label{second_order_part}
\frac{\Delta\nu}{\nu_0}\approx\frac{1}{2}\frac{\delta\ddot{\nu}(t)}{\nu_0}\left(\frac{T_{{\mathrm{samp}}}}{2}\right)^2,
\end{equation}
where $T_{\mathrm{samp}}$-- length of sub-interval, i.e. one year;
on the other hand, we can detect fractional deviations of
frequency $\frac{\Delta\nu}{\nu}$ at $R=10^{-14}$ level.

  Assuming flat spectrum of GWB
($\Omega_{gw}(k)=const$) and using (\ref{nuddotsquaremean3}) we
arrive at:
\begin{equation}
\label{detection}
\Omega_{gw}<\frac{32}{\pi^2}R^2\left(\frac{T_{\mathrm{H}}}{T_{\mathrm{samp}}}\right)^2,
 \end{equation}
where $T_{\mathrm{H}}\equiv\frac{1}{H_0}$ is Hubble time;
detection limit is $\Omega_{gw}\sim 10^{-7}$. This estimation  is
certainly too superficial and will be reconsidered with use of
real pulsar array data.

\section{Conclusions}
Measuring rotational parameters of pulsar can provide  valuable
information about processes with characteristic times larger than
time span of pulsar observations. One of the most interesting kind
of these processes is  influence of ultra long-wavelength GWB on
pulsar timing. The second derivative of rotational frequency can
be used to put limits on the density of GWB in the frequency
region $10^{-12}$ to \hbox{$10^{-8}$\,Hz}. The limits for
$\Omega_{gw}$ that come from the suggested method:
$$\Omega_{gw}<2\times10^{-6}.$$
 Also, the time series of the frequency $\nu$ from different
pulsars in pulsar array can be cross-correlated pairwise.
Comparison of  angular dependence of obtained functions with known
sample of GWB induced correlation similarly to usual pulsar timing
residual analysis makes possible to detect GWB in the frequency
range of our interest. That method can probe GWB down to energy
density $\Omega_{gw} < 10^{-7}$.

\section*{Acknowledgments}
I would like to thank V.A.~Potapov and D.~Baskaran  for useful
discussions and fruitful suggestions.  Also I wish to thank the
anonymous referee for his comments that helped  to improve the
paper. This work was supported by RFBR Grants No. 09-02-00922-a
and No. 07-02-01034-a. This research has made use of NASA's
Astrophysics Data System.

%\appendix
%
%\section[]{}
%
%
%
%\bsp

\label{lastpage}

\end{document}